# CIX – A Detector for Spectral Enhanced X-ray Imaging by Simultaneous Counting and Integrating

H. Krüger*[a], J. Fink[a], E. Kraft[a], N. Wermes[a], P. Fischer[b], I. Peric[b], C. Herrmann[c], M. Overdick[c], W. Rütten[c]

[a]University of Bonn, Physics Department, Nussallee 12, 53115 Bonn, Germany;
[b]University of Mannheim, Institute of Computer Engineering, B6, 26, 68131 Mannheim, Germany;
[c]Philips Research Europe, Weisshausstrasse 2, 52066, Aachen, Germany

## ABSTRACT

A hybrid pixel detector based on the concept of simultaneous charge integration and photon counting will be presented. The second generation of a counting and integrating X-ray prototype CMOS chip (CIX) has been operated with different direct converting sensor materials (CdZnTe and CdTe) bump bonded to its 8x8 pixel matrix. Photon counting devices give excellent results for low to medium X-ray fluxes but saturate at high rates while charge integration allows the detection of very high fluxes but is limited at low rates by the finite signal to noise ratio. The combination of both signal processing concepts therefore extends the resolvable dynamic range of the X-ray detector. In addition, for a large region of the dynamic range, where counter and integrator operate simultaneously, the mean energy of the detected X-ray spectrum can be calculated. This spectral information can be used to enhance the contrast of the X-ray image. The advantages of the counting and integrating signal processing concept and the performance of the imaging system will be reviewed. The properties of the system with respect to dynamic range and sensor response will be discussed and examples of imaging with additional spectral information will be presented.

## 1. INTRODUCTION

Imaging by counting individual X-ray photons has become feasible with the advance in CMOS micro-electronics and the availability of fast direct converting sensor materials like CdTe and CdZnTe. The concept of hybrid pixel detectors, in which the pixels of a sensor crystal are connected to individual electronic cells of an ASIC, was largely driven in the past by the detector development for particle physics experiments. In this context the first photon counting ASICs for X-ray imaging applications where developed [1], [2]. By processing the signal of individual absorbed X-ray photons, it is in principle possible to measure the energy spectrum of the absorbed quanta per pixel, which allows for example the calculation of the object dependent beam hardening. Although there are developments, which in fact pursue a per pixel ADC [3], constraints given by the available layout area for the pixel electronics, power consumption and count rate favor a more basic approach. Dual- and multi-threshold counting schemes were proposed to access spectroscopic information of the absorbed signals [4], [5], [6]. However, a fundamental limitation of photon counting concepts lies in the practical bandwidth of the analog signal processing and the therefore limited maximum count rate per pixel which is in the order of a few MHz [5], [8]. In addition the pile-up effect leads to a smearing of the effective counter thresholds and deteriorates the energy resolution.

The concept of the CIX chip follows another approach to access the spectral energy information. By counting individual photons and simultaneously integrating their deposited charge, the mean energy of the absorbed photons can be calculated. Of course this energy reconstruction is only feasible in a range in where the dynamic ranges of the photon counter and integrator overlap. The lower range is limited by the noise performance of the integrator and the upper limit is given by the maximum photon count rate. Yet even beyond the saturation point of the photon counter, the simultaneous operation of the integrator extends the detectable photon flux by an order of magnitude.

The implementation of the CIX signal processing concept will by explained in the next chapter and the characterization of the photon counter and the integrator will be presented in chapter 3. Measurements with bump-bonded CdTe and CdZnTe sensor crystals will be presented in chapter 4. The electrical performance of the CIX connected to a sensor will be revised and characteristic issues of the sensor materials will be addressed.

## 2. IMPLEMENTATION

The CIX detector consists of a pixelated CMOS signal processing chip to which a direct converting sensor crystal is bump bonded. The basic function blocks in each pixel cell of the CMOS chip are: a charge sensitive amplifier (CSA), a current integrator, a feedback circuit and different counters for the CSA and the integrator (fig. 1). The charge signal generated by an absorbed X-ray photon is amplified by the CSA which triggers the photon counter when the programmed threshold has been exceeded. The feedback circuit not only discharges the CSA's feedback capacitance $C_F$ but also mirrors the feedback current into the integrator. Due to this feedback current replication the integrator is able to measure the total accumulated charge for a given time interval. The integrator implementation is similar to a current to frequency converter and the value of the measured charge and the corresponding time interval is stored as a binary number in the pixel counters.

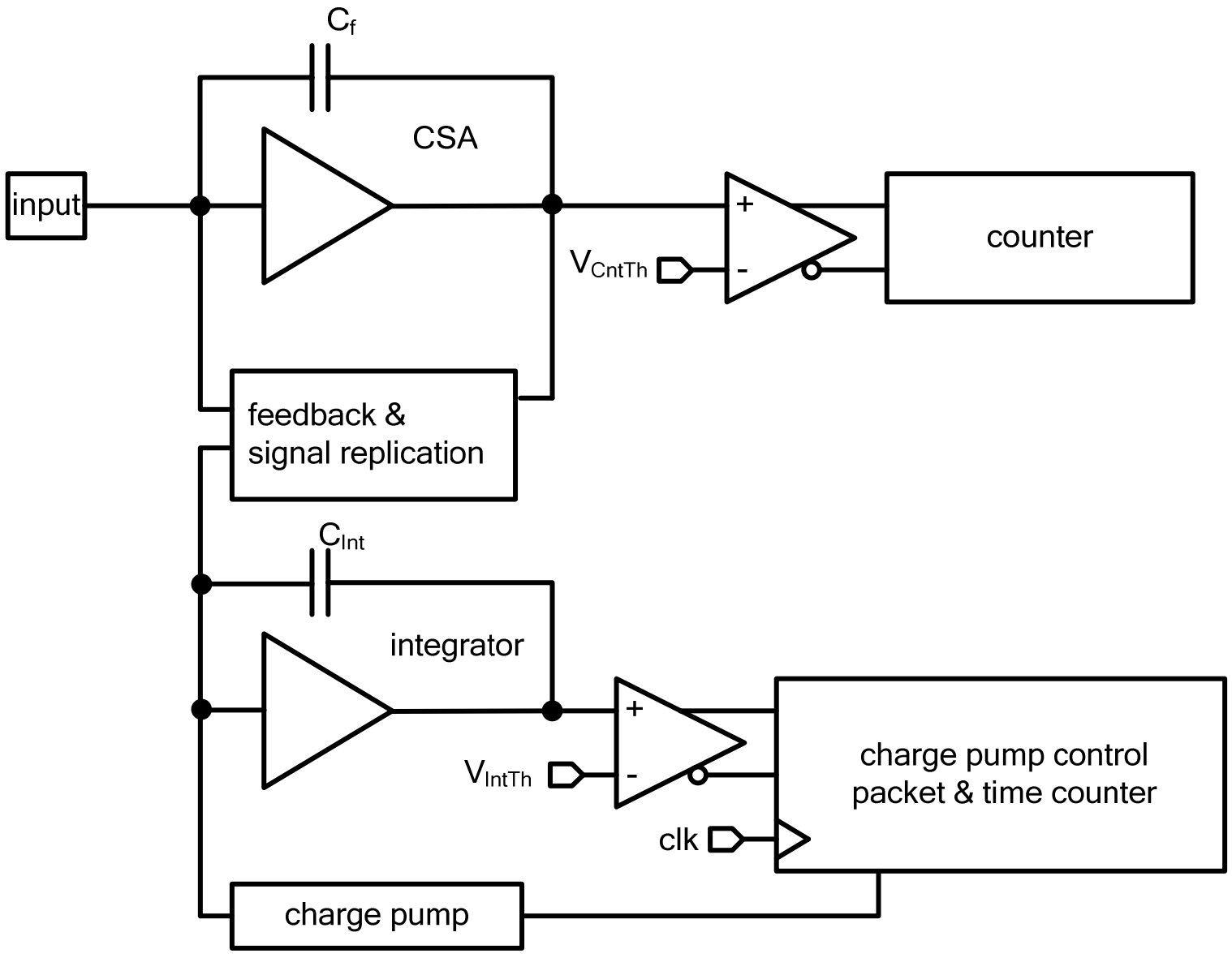

Fig. 1: Basic function blocks of the CIX pixel cell. The upper part shows the components of a conventional photon counting scheme: a charge sensitive amplifier, a comparator with a global threshold voltage, local programmable tuning (not shown) and a counter. The lower part depicts the components of an integrator which is implemented as a current to frequency converter comprising an integrating amplifier, a comparator, a switched charge pump and a digital part with control logic and counters. The input current to the integrator is a replication of the charge sensitive amplifier's feedback current. The pixel cell input on the left is connected to the sensor. Alternatively, the input can be driven by different on-chip charge injection circuits during test and calibration.

### 2.1 Photon Counter and Feedback with Signal Current Replication

The CSA of the photon counter is implemented with a differential input amplifier (AMP1) and two fully differential transconductance amplifiers (OTA1 and OTA2) in the feedback path (fig. 2). A two stage comparator with a programmable global threshold connects the output of the CSA to a 16 bit counter. To reduce the threshold dispersion introduced by the intrinsic CMOS device mismatch, a 6 bit DAC in each pixel adjusts the effective local threshold individually. The switches A, B, C and D enable different operation modes of the feedback circuit. In favor of a more comprehensive explanation of the circuit's functionality, the function of OTA2 shall first be omitted. In an equilibrium state when no input signal is present and the feedback capacitance $C_f$ is discharged, OTA1's output current is zero and the output voltage of AMP1 is equal to $V_{CountBaseLine}$. Once for example a negative current pulse is applied to the input, the resulting charge is integrated on $C_f$ and the output voltage of AMP1 rises. At the same time, the non-inverting output of OTA1 begins to deliver current to the input node which eventually compensates the integrated charge until the equilibrium state is reached again. The inverting output of OTA1 delivers the same amount of feedback current into the input of the integrator via the switch A. As a result the integrator obtains an exact copy of the original CSA input charge.

The purpose of OTA2 is the compensation of static offset currents arising from sensor leakage and device offsets in OTA1. This offset current would otherwise result in a shift of the CSA output baseline and an offset in the integrator measurement. There are different modes of offset current compensation available: In the so called *static leakage current compensation mode*, the offset current is sampled in one phase, during which no input signal is present. In this sample phase, the integrator is disconnected from the feedback (A and B open) and switch D at the inverting input of OTA2 is closed. The potential of the sampling capacitor $C_S$ will settle to a value such that the differential input voltage of OTA1 becomes zero (i.e. AMP1 output equal to $V_{countBaseLine}$). In return, this demands that no current flows into the CSA input. Consequently any input current is compensated by the non-inverting output of OTA2. Once this state is reached, switch D is opened (hold phase) and the integrator is again connected to the feedback by closing switch A. By means of the other switches B, C and D and a third OTA which is not shown in fig. 2, other modes of feedback operation and leakage current compensation can be selected. However their description is beyond the scope of this paper. A detailed description of the feedback and its different operation modes can be found in [7] and [9].

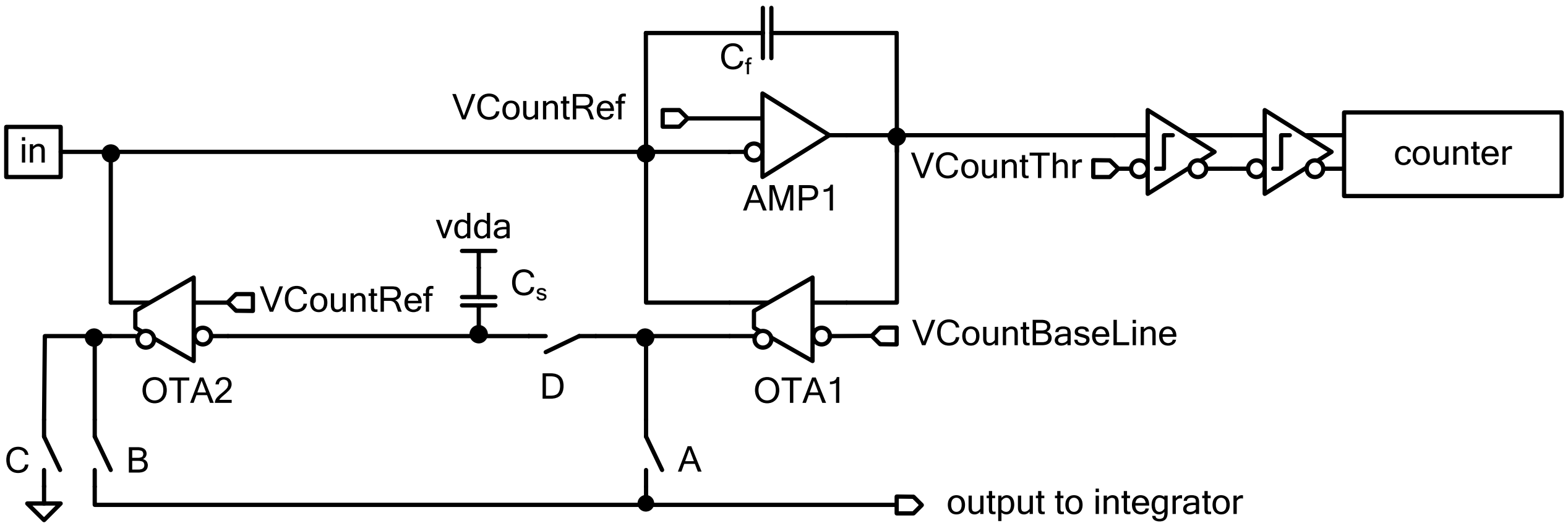

Fig. 2: More detailed view of the CSA and the circuit used to provide feedback and the signal replication for the integrator. See text for explanation of the circuit.

### 2.2 Integrator

The integrator is implemented similar to the architecture described in [10]. The underlying concept is a first order sigma-delta modulator which is often used for high precision measurements. As shown in figure 3, the integrator output is connected to a two stage comparator, while its input is connected to a charge pump. A negative current at the input charges $C_{Int}$ so that the output will eventually exceed the comparator threshold. This triggers the control logic which activates the charge pump so that $C_{Int}$ will be discharged with the charge packet $Q_{pkt}$. As long as the input signal is present, $C_{Int}$ will undergo subsequent charge- and discharge cycles. The output voltage of the integrator will hence show a characteristic saw-tooth-like waveform. Every time a charge packet is subtracted from the input, the charge packet counter is incremented. To calculate the mean current the charge packet count $N_{pkt}$ has to be divided by the time $\Delta t$. This interval is given by the time between the first and the last pump event and is measured in terms of clock cycles $N_t$.

$$I_{Signal} = \frac{N_{pkt} \cdot Q_{pkt}}{\Delta t} = \frac{N_{pkt} \cdot Q_{pkt} \cdot f_{clk}}{N_t} \tag{1}$$

Therefore the comparator must at least trigger two pump events to produce a valid measurement. This sets a systematical lower limit on the input current that the integrator can measure. An additional current source $I_{IntBias}$ can be used to inject a programmable offset current into the integrator, such that even without sufficient input current, at least two pump events are present in every measurement interval (frame time). The maximum detectable input current is set by the system clock frequency $f_{clk}$ and the charge packet size $Q_{pkt}$. Note that minimum and maximum current limits are both proportional to the charge packet size. The dynamic range of the integrator is given by the frame time $t_{frame}$ and the system clock frequency. For example a clock rate of 20 MHz and a frame time of 2 ms lead to a dynamic range of

$$f_{clk} \cdot t_{frame} = 20\ MHz \cdot 2\ ms = 40{,}000 \rightarrow 15.3\ \text{bits.} \tag{2}$$

The dispersion of the comparator thresholds is not a concern for the operation of the integrator, since it will lead only to an offset in the output baseline. The precision of a measurement – and the variation between pixels – is determined by the quality and the matching of the charge pumps only.

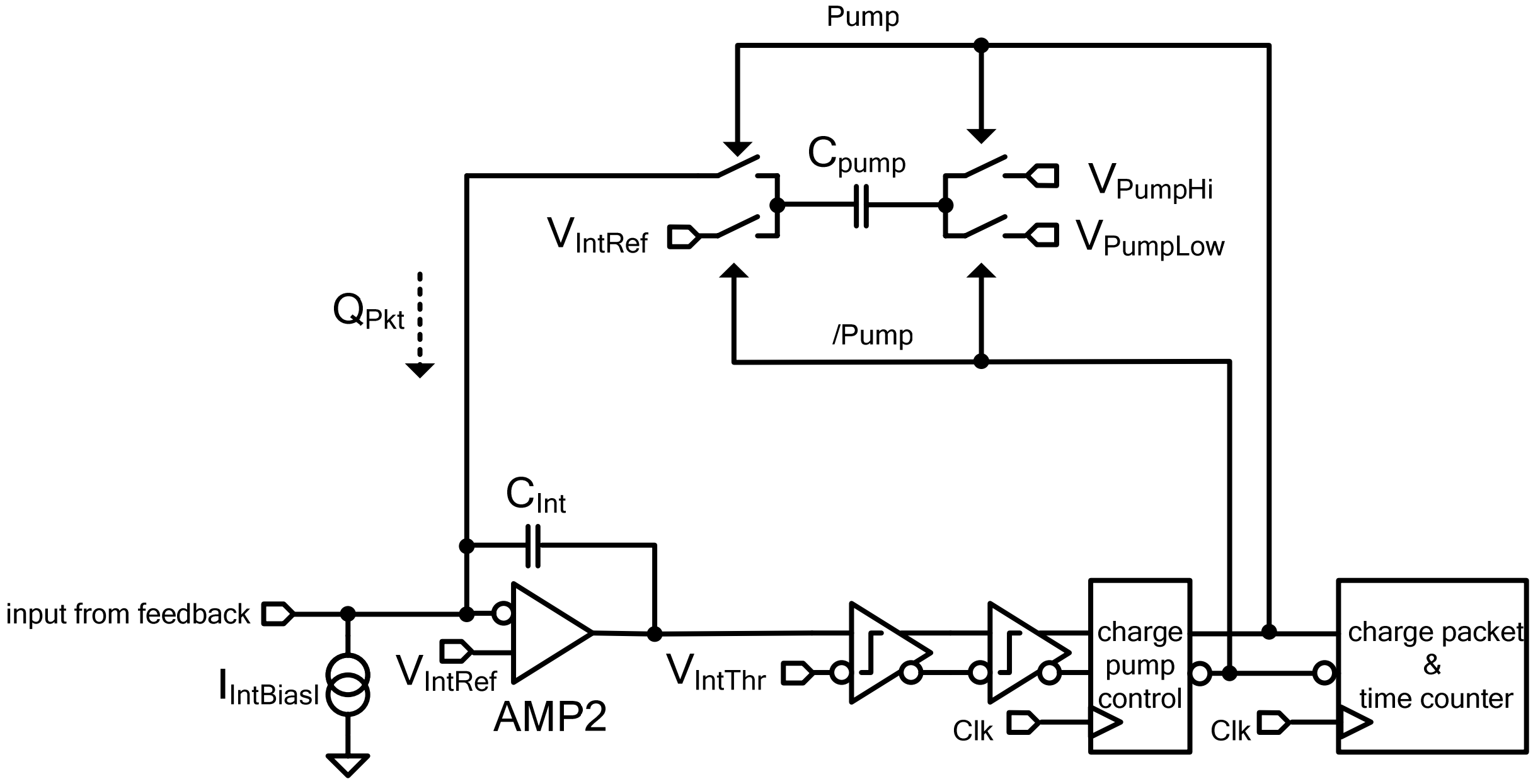

Fig. 3: Implementation of the integrator. The output of the integrator AMP2 is discriminated by a double stage comparator which triggers the switched capacitor charge pump around $C_{pump}$ synchronized to the system clock. The number of pump cycles and additional timing information is stored in16 bit wide counters and latches, respectively.

The charge pump shown in figure 3 is implemented as a switched capacitor charge pump. When the comparator output is inactive, the capacitor $C_{pump}$ is pre-charged: One electrode is connected to a reference voltage $V_{IntRef}$, which is equal to the input voltage of the integrator, while the other electrode is connected to $V_{PumpLow}$. When the comparator triggers the charge pump, $C_{Pump}$ is connected between the integrator input and $V_{PumpHi}$. The control logic ensures at least one pre-charge cycle after pump event. By variation of the bias voltages $V_{PumpHi}$ and $V_{PumpLow}$ the transferred charge packet size $\Delta Q$ can be adjusted:

$$\Delta Q_{pkt} = C_{Pump} \cdot (V_{PumpHi} - V_{PumpLow}) \tag{3}$$

Since capacitors implemented in CMOS technologies show a good matching, the dispersion of the charge pumps is very small. The current version of the CIX chip also features alternative charge pump circuits [9], [11].

### 2.3 Read-out Architecture

The cells of the pixel matrix are organized in columns, which provide a 16 bit wide bus to transfer the data of the pixel counters to the end of the column. There the data is serialized and transferred via LVDS outputs to the data acquisition system. Due to the use of dedicated latches for the pixel counters and double buffering in the end of column logic, the CIX chip can be read out during data acquisition, enabling an almost dead-time free operation. The extensive use of current-mode logic guarantees that no additional crosstalk from the activity in the readout circuitry compromises the low noise analog performance [9], [12]. However, this advantage is paid for with increased static power dissipation in the digital circuits. This leads to a power consumption of 3.2 mW per pixel of which 98 % is caused by the digital cells.

The current prototype chip CIX 0.2 was implemented in a 0.35 µ CMOS technology and has a matrix layout of 8 x 8 pixels with a pixel size of 500 µm x 250 µm. The die size is 5.6 mm x 4.1 mm.

## 3. MEASUREMENTS

Before the ASIC was bump bonded to a sensor crystal, the performance of the pixel electronics was measured without sensor by using on-chip charge injection circuits. In particular the dynamic range and the electronic noise of photon counter and integrator have been evaluated with this method.

### 3.1 Performance of the Photon Counter

The photon counter was characterized using an on chip charge injection circuit. Table 1 summarizes the key figures of the photon counter. The numbers given are valid for typical bias settings with leakage current compensation switched off. Active leakage current compensation increases the noise by about 1 electron per nA. Another important aspect of a photon counter is the minimum threshold. Figure 4 shows the noise induced count rate versus the threshold setting. The noise count rate lies below 1 kHz for threshold settings above 6 mV which is equivalent to 387 $e^-$. Above 9 mV (581 $e^-$) the count rate is below one count per minute. This plot also shows that there is no significant digital-to-analog crosstalk, which would otherwise distort the Gaussian-shaped curve of the measured noise count rate.

Table 1. Features of the photon counter. Depending on the feedback bias and threshold settings, the typical maximum count rate varies between 6 and 20 MHz (equidistant pulses).

| | |
|---|---|
| equivalent noise charge | 119 $e^-$ |
| noise excess per 100 fF input cap. | 37.5 $e^-$ |
| feedback capacitance $C_f$ | 10.35 fF |
| max. count rate (typ., 2 fC pulses) | 12 MHz |
| threshold dispersion (adjusted) | 36 $e^-$ |

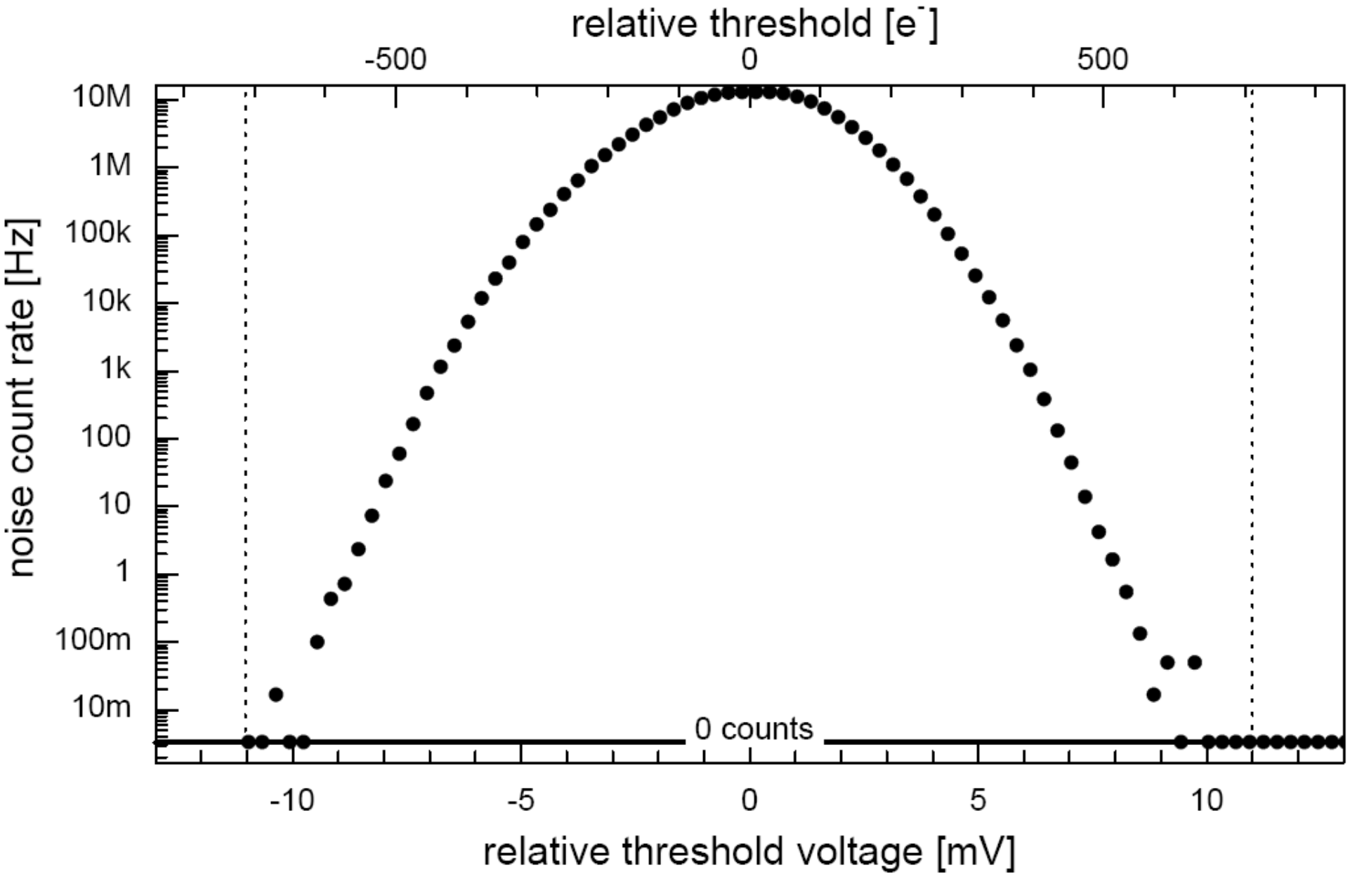

Fig. 4: Measurement of the noise count rate (logarithmic scale) [9]. The Gaussian-shaped curve of the measured noise count rate demonstrates that there is no significant digital-to-analog crosstalk, which would lead to distortions or hysteresis in the measurement.

### 3.2 Performance of the Integrator

The characterization of the stand-alone performance of the integrator was done by direct injection of signals into the integrator, bypassing the feedback network. An additional offset current of 770 pA was injected via $I_{IntBias}$ (see figure 3) in order to extend the dynamic range to the lowest possible value. Figure 5 shows the dynamic range of the integrator.

The measurement is in good agreement with the theoretical value of 4.6 decades according to the parameters given in equation 2. At the upper end the integrator saturates at 200 nA as expected due to the given settings of $f_{clk}$ = 20 MHz, $Q_{Pkt}$ = 10 fC and a frame time of 2 ms. The lower end shows fluctuations due to the discrete nature of the injected signals.

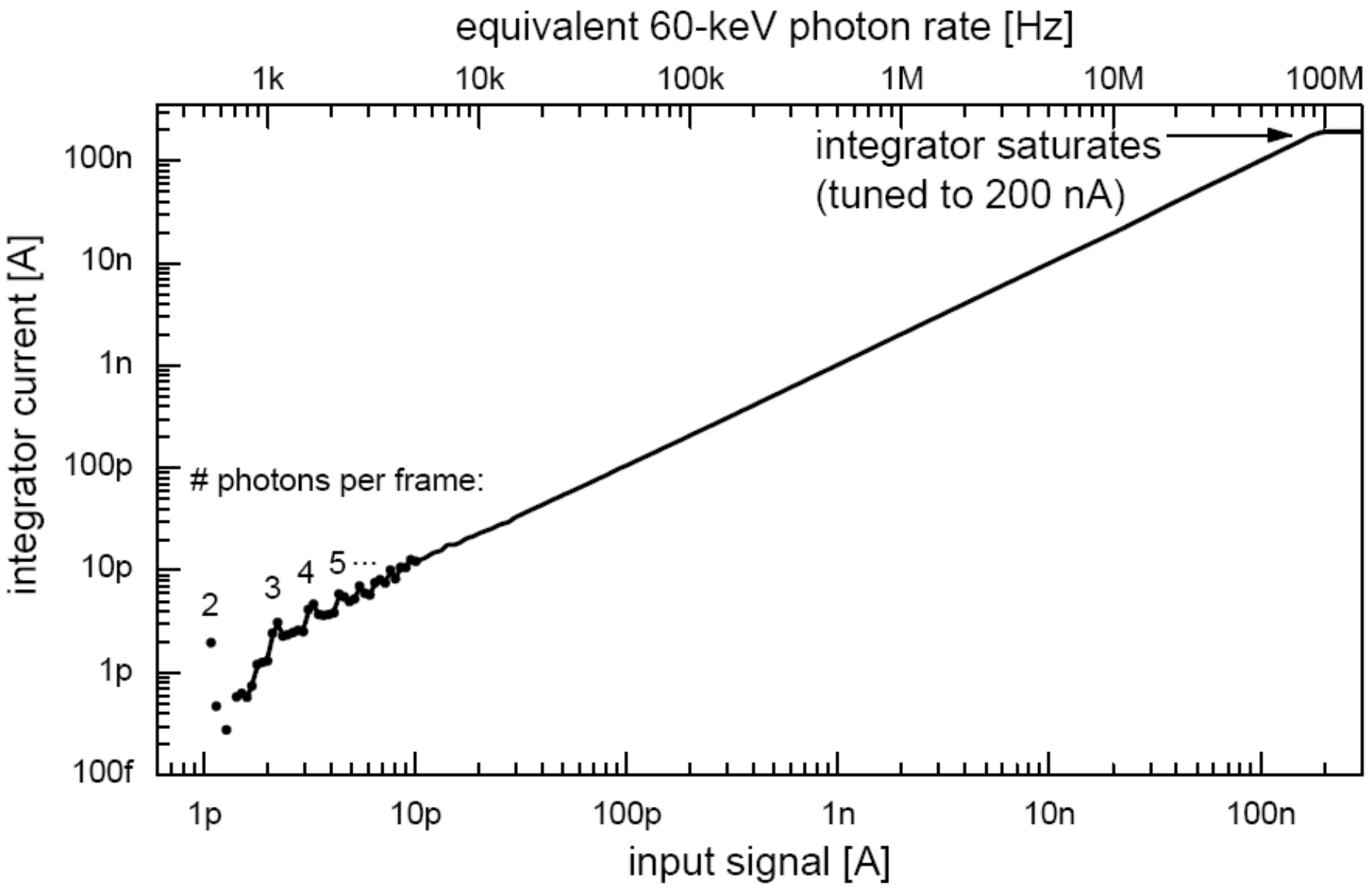

Fig. 5: Dynamic range of the integrator: With a system clock of 20 MHz and a frame time of 2 ms, an input signal range from 2 pA up to 200 nA can be achieved. A pulsed current source (2.1 fC pulse size, equidistant) generated the signal up to 30 nA input current. Above that value a continuous current source was used [9].

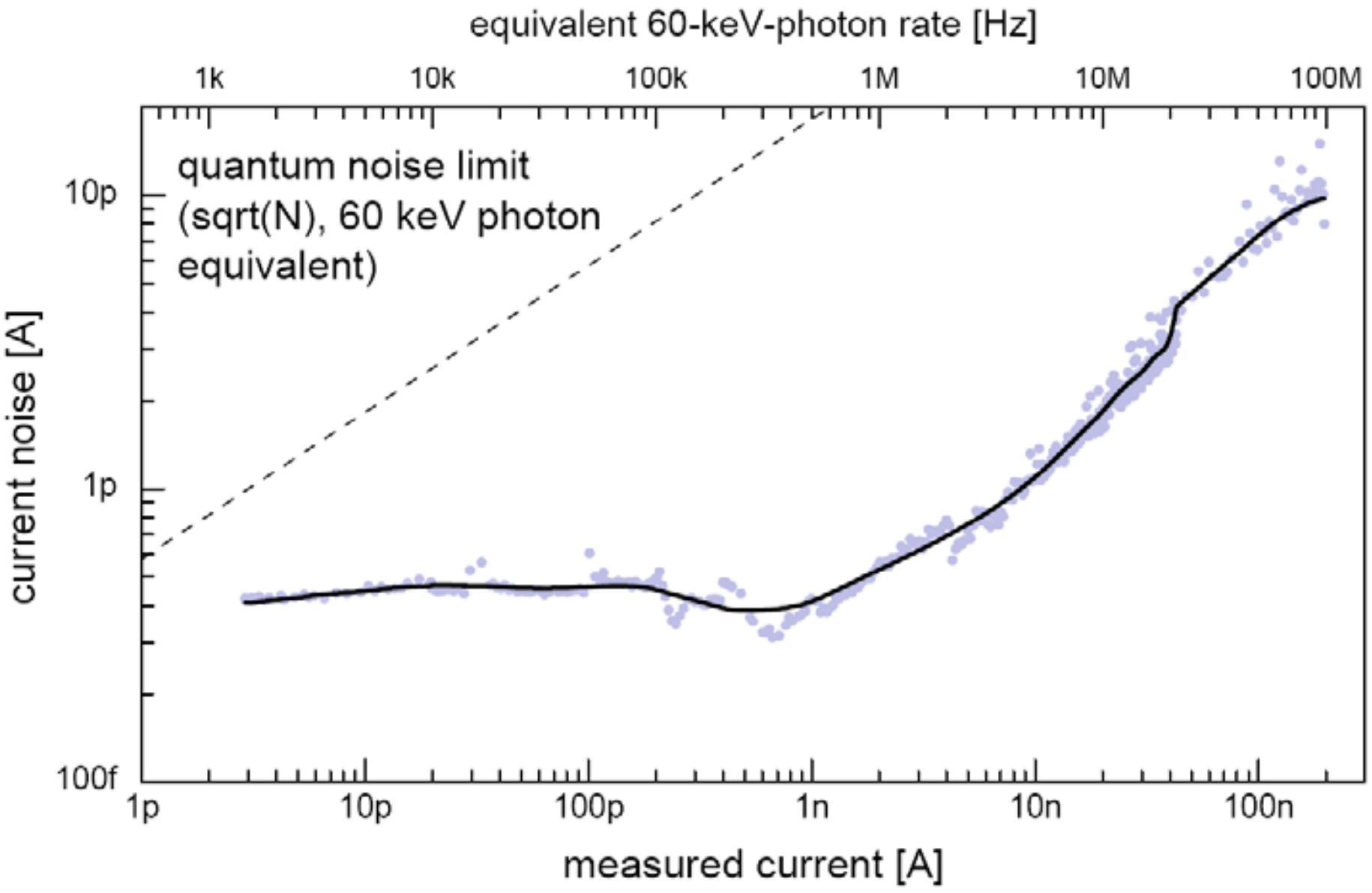

Fig. 6: Absolute value of the integrator noise for a frame time of 6 ms. The dashed line shows the expected quantum noise from random 60 keV photon equivalent signals [9].

The integrator noise was measured by calculating the standard deviation of 100 subsequent measurements. Figure 6 shows the measured noise as a function of the signal current. Up to 1 nA, the noise is almost constant at about 0.4 pA. This noise floor is induced by injecting an additional offset current of 770 pA in order to extend the lower limit of the dynamic range. At the maximum input current of 200 nA, the noise has increased to about 10 pA which corresponds to a signal-to-noise ratio of 20,000/1. Note that even at the lowest input signals, the noise is well below the statistical limit introduced by the quantum fluctuations that a real X-ray signal would exhibit.

### 3.3 Simultaneous Counting and Integration

For the characterization of the chip operating in simultaneous counting and integrating mode, the feedback circuit was set to *static leakage current compensation* (see 2.1). Figure 7 shows the response of the photon counter and the integrator to the injected input signal. While the photon counter measures photon rates up to 12 MHz (corresponding to 24 nA input current) the current integrator increases the systems dynamic limit to 200 nA (corresponding to 95 MHz of 60 keV photon equivalent pulses). At the lower end, the integrators noise contribution becomes noticeable for currents smaller than 100 pA (one-standard-deviation noise interval in fig. 7). Note that the noise in the current measurement has increased from 0.4 pA to 24 pA. This discrepancy compared to the stand-alone operation is caused mainly by additional kT/C noise induced by sampling of the leakage current. Above 250 pA however, the signal-to-noise ratio of the current measurement is better than 10. As a result, the performance of the CIX allows the reconstruction of the mean photon energy within an input range of two decades, from 250 pA to approximately 25 nA.

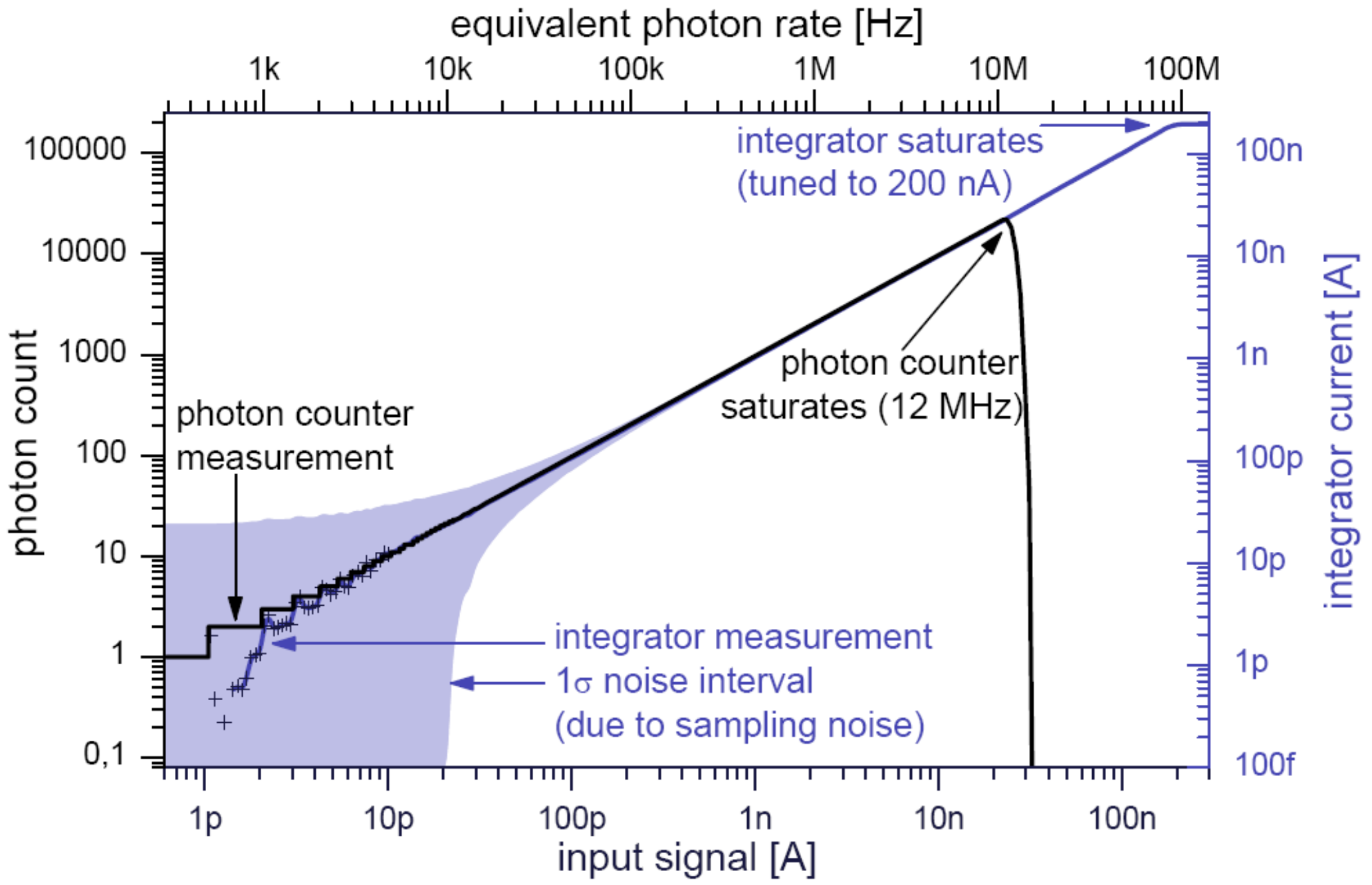

Fig. 7: Measurement of the dynamic range and noise for simultaneous counting and integrating. A pulsed current source (2.1 fC pulse size, equidistant) generated the input signal up to 30 nA. Above that value, a continuous current source was used. Other settings were: 2 ms frame time, 20 MHz clock frequency, 10 fC charge packet size, 1.12 fC photon counter threshold [9].

## 4. SENSOR CHARACTERIZATION

The CIX chip was bump-bonded to different CdTe and CdZnTe sensor crystals which were metalized with 8 x 8 pixelated electrodes matched to the dimensions of the CIX pixel cells (250 µm x 500 µm). The CdZnTe crystals (eV Products) are 3 mm thick and the CdTe crystals (Acrorad) are 1mm thick. Both crystal types were metalized with Pt electrodes. The bump bonding was done with gold-studs and non-conductive glue.

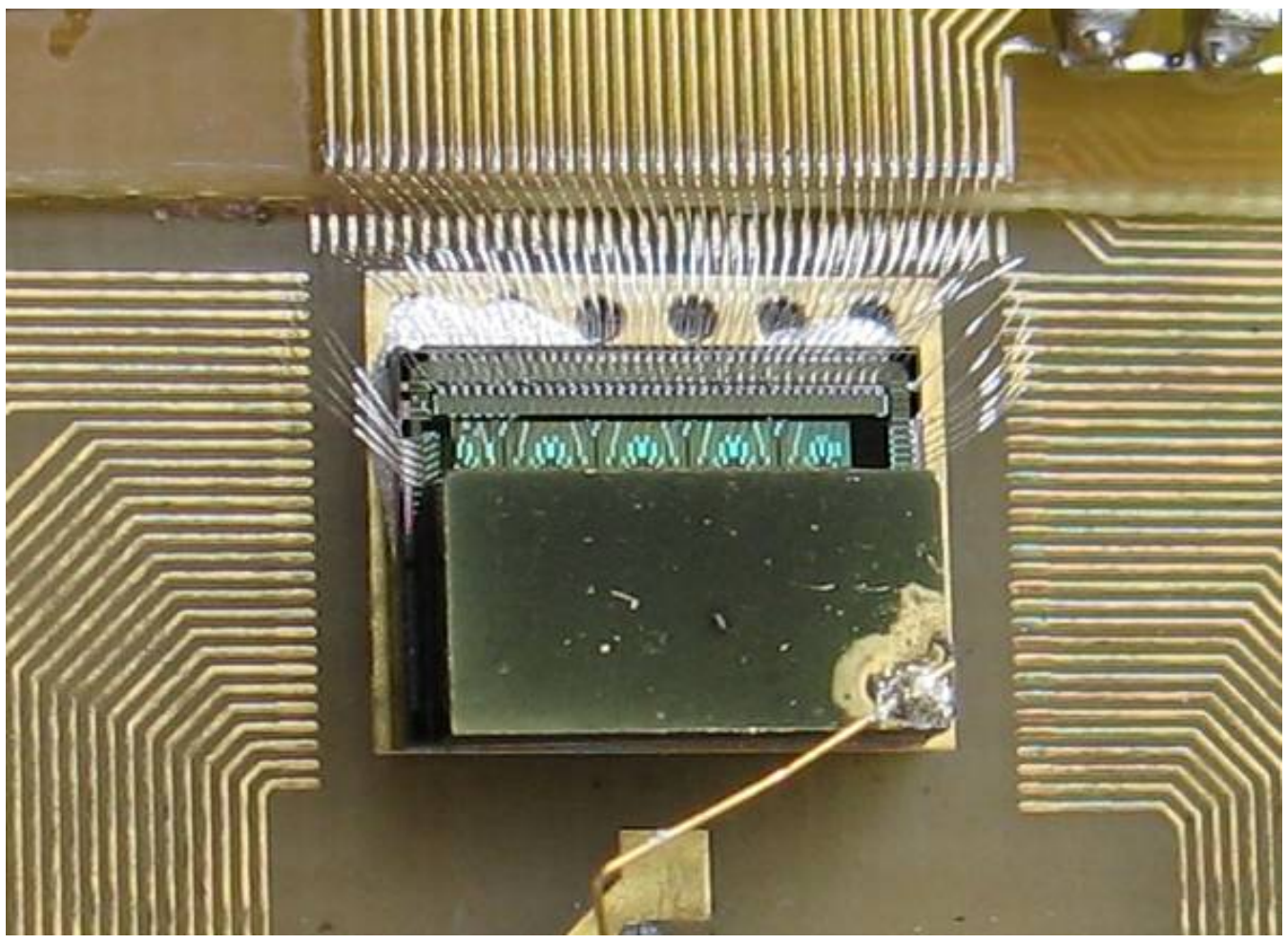

Fig. 8: Photograph of the CIX 0.2 prototype chip bump bonded to a 1 mm thick CdTe sensor crystal. In the lower right corner of the metalized detector backplane the bias contact is made. The chip extends under the sensor area at the top where the wire bonds connect the chip periphery to the data acquisition system.

A first investigation studied the influence of the capacitive load from the connected sensor crystal. The ENC of the photon counter has increased form 119 $e^-$ to 333 $e^-$ which corresponds to an input capacitance of about 0.5 pF (see table 1). For the measurement shown in figure 9, the integrator was directly connected to the CdTe sensor (see chapter 3.2) and an on-chip current source was used to inject the signal. In this case no additional offset current was applied such that the lower integrator limit is somewhat higher compared to the plot in figure 6. The measured noise for 3 ms frame times is an order of magnitude below the quantum limit. Even for 200 µs frame times, the error of the current measurement would still be dominated by the quantum fluctuations.

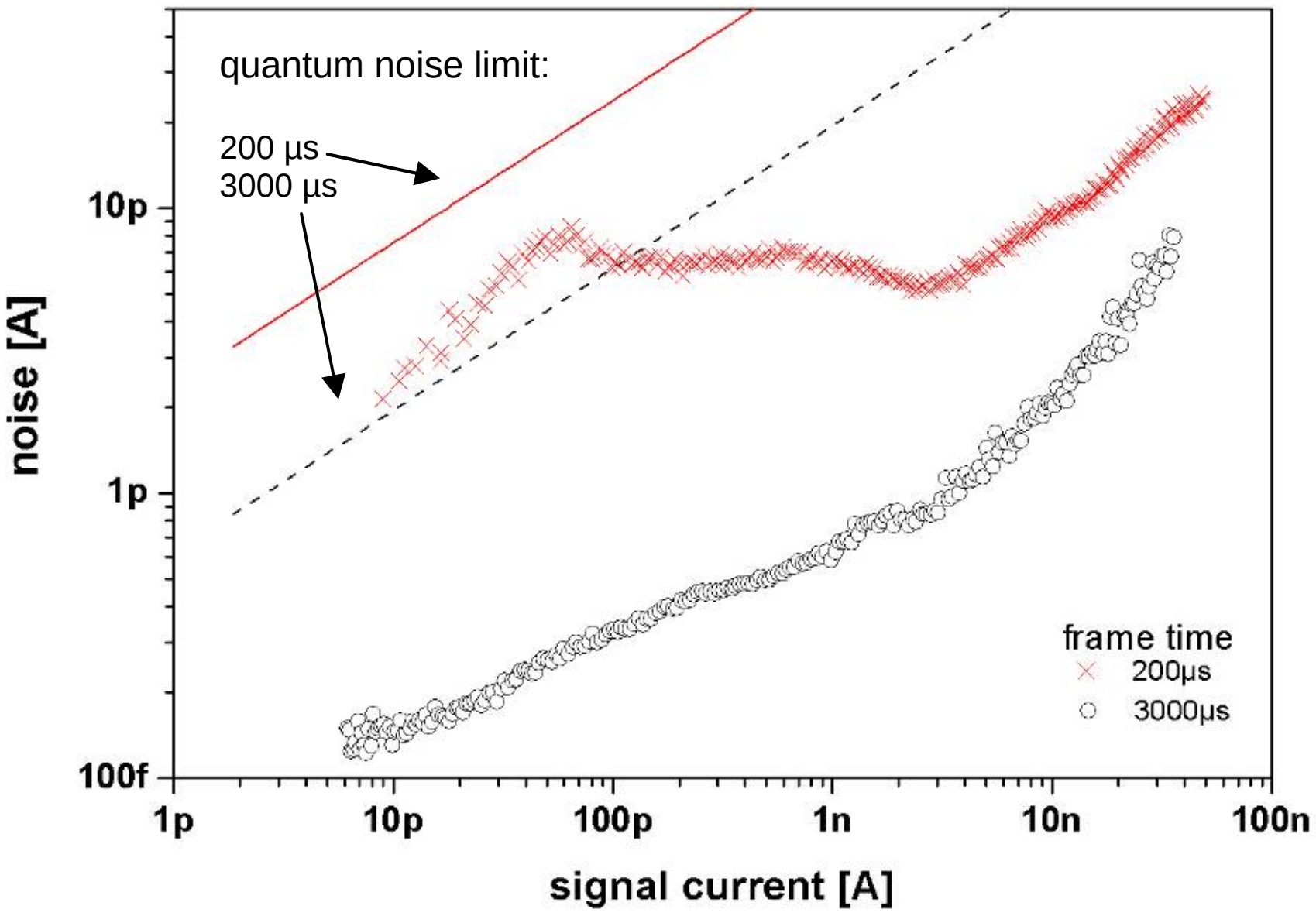

Fig. 9: Integrator noise with a 1 mm CdTe sensor connected to the chip (stand-alone operation of the integrator, direct injection, input signal generated by on-chip current source, detector not biased).

For measurements with real X-ray signals the detector was exposed to the beam of a micro-focus X-ray tube with 90 kVp. Figure 10 shows the measured integrator current and photon count rate as a function of the tube current (mean across the detector). Compared to the measurement without sensor (figure 7) the (measured) maximum count rate has decreased to about 2.5 Mcps, while about 4.4 Mcps would be expected when taking Poissonian (rather than equidistant) arrivals into account. The mechanism causing this deviation is not fully understood yet. One reason is the stronger influence of the pile-up effect due to the polychromatic nature of the X-ray source. A second contribution is the detector leakage current which lowers the effective feedback current unless compensated. As explained below, the leakage compensation was deactivated in this measurement. This caused longer decay times of the CSA pulses, further increasing the pile-up. Additional contributions arise from the finite charge collection time and the capacitive load of the sensor.

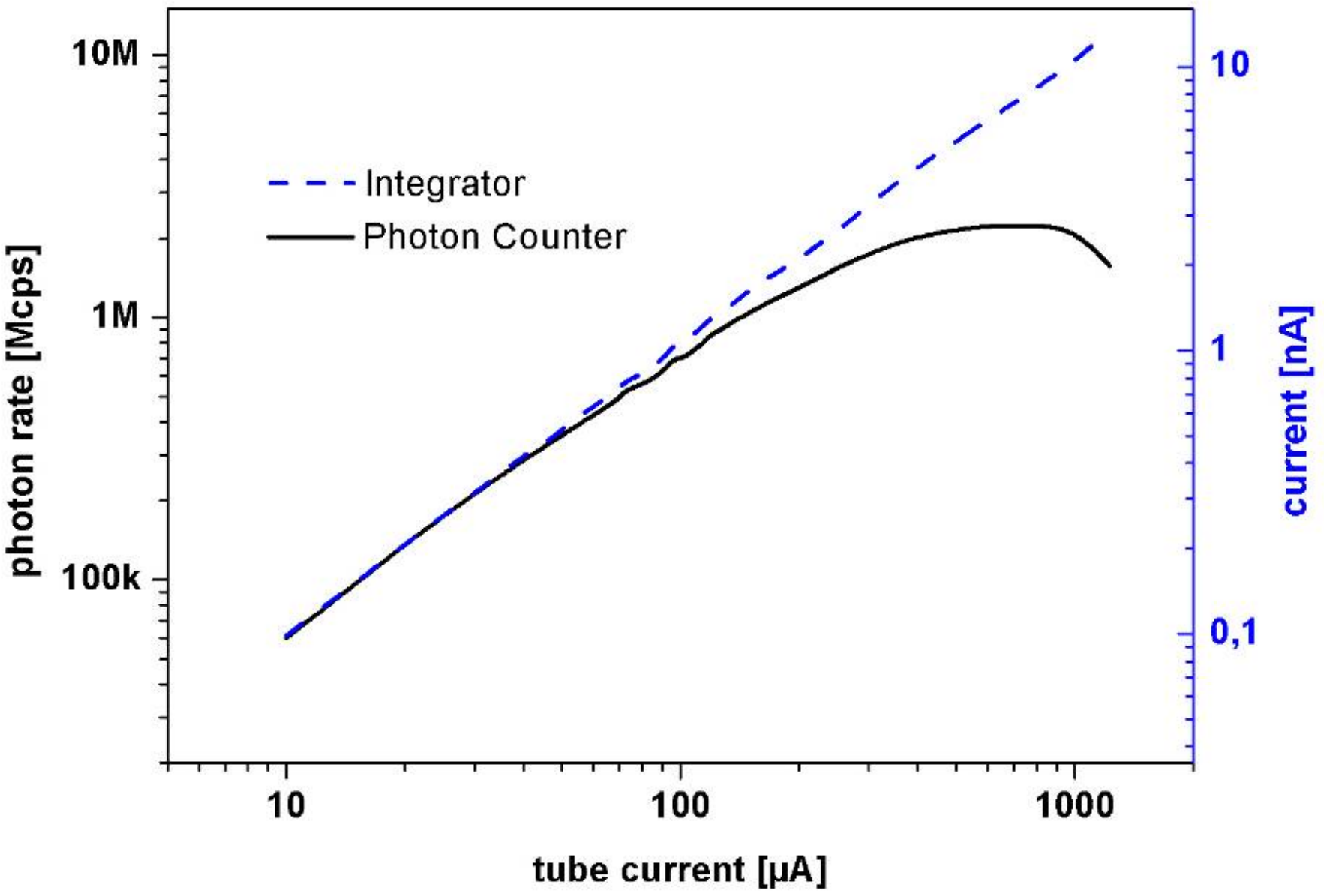

Fig. 10: Measurement of the dynamic range with the 3 mm CdZnTe sensor (mean value of all pixels, 90 kVp X-ray tube, 150 µm Be and 300 µm Al filter). The dashed line shows the integrator current, the straight line shows the photon count rate. The measurement range of the integrator is limited by the maximum photon flux of the micro-focus X-ray tube.

### 4.1 Evaluation of the Spectroscopic Performance

The evaluation of the sensor crystals spectral response revealed unexpected difficulties. The leakage current of the CdZnTe and CdTe crystals shows a temporal variation and a strong dispersion between the pixels (table 2). As described in section 2.1, the leakage current compensation circuit is designed to compensate such fluctuations by sampling the leakage current in the gaps of a pulsed X-ray beam. Unfortunately our X-ray source is unable to provide such gaps with the required timing precision. As a result, the measurements with the X-ray tube cannot be compensated for the leakage current at the time of writing.

Table 2: Mean leakage current per pixel and one-sigma standard deviation across the 8 x 8 pixel matrix, measured at approximately 40°C.

| | CdTe (1 mm) | CZT (3 mm) |
|---|---|---|
| bias voltage | -400V | -1000V |
| mean leakage current | 5.1 nA | 0.39 nA |
| dispersion | 1.1 nA | 0.14 nA |

This dispersion of the leakage current translates into an inhomogeneity of the count rate via their influence on the effective feedback current. The left plot in figure 11 gives an example of the measured count rates when illuminating a CdZnTe sensor with a homogeneous X-ray beam. On the right side of that figure the measured current is shown. There is a noticeable correlation between count rate and current measurement for most pixels. Studies to quantify these relations between count rate, leakage current and sensitivity are currently under way.

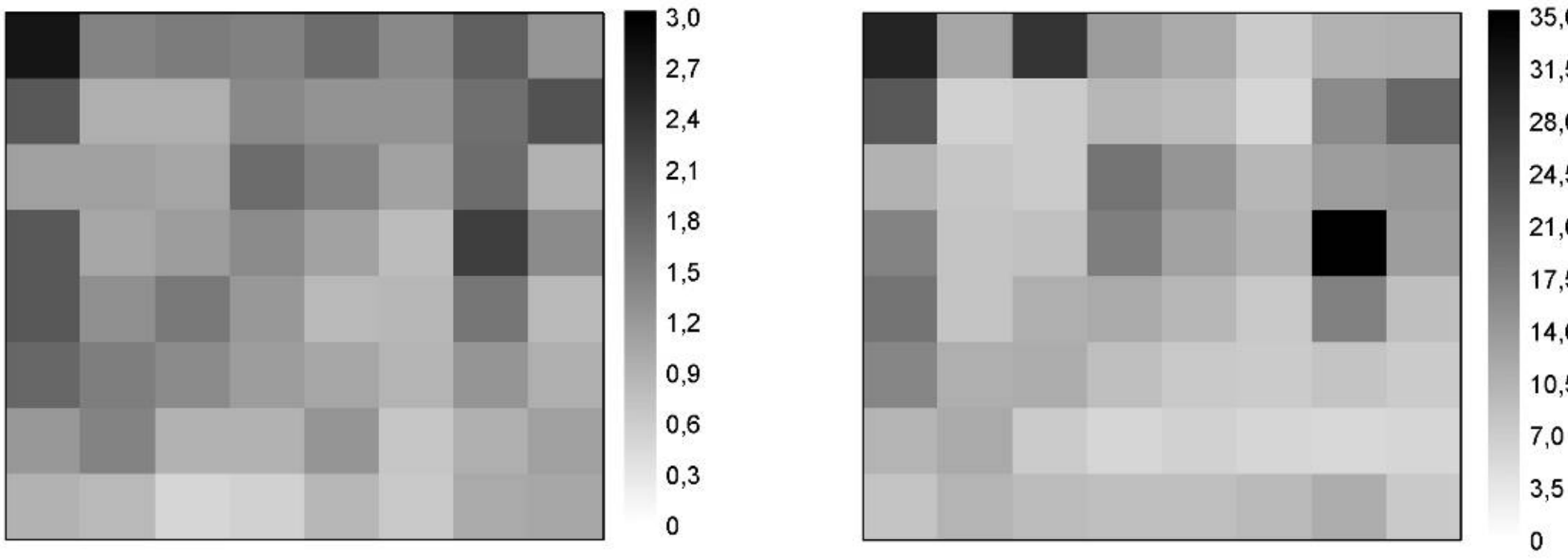

Fig. 11: Response of the CdZnTe sensor to a homogeneous 90 kVp X-ray beam: counter values in Mcps (left) and integrator values in nA (right image). The tube current was set to 200 µA to operate the counters below saturation (see figure 10).

The response of CdZnTe to the manually switched X-ray tube is shown in figure 12. As expected, most of the pixels show the typical afterglow effect [13], which is revealed by an exponentially decaying signal with a time constant in the order of minutes. Surprisingly, however, many pixels show the opposite behavior: after the tube was switched on, their signal current showed a peak followed by an exponential decay (not shown in Fig. 12). This coincided with a negative 'under-shoot' when the tube was switched off (cf. Fig. 12). The reason for this behavior is still unclear.

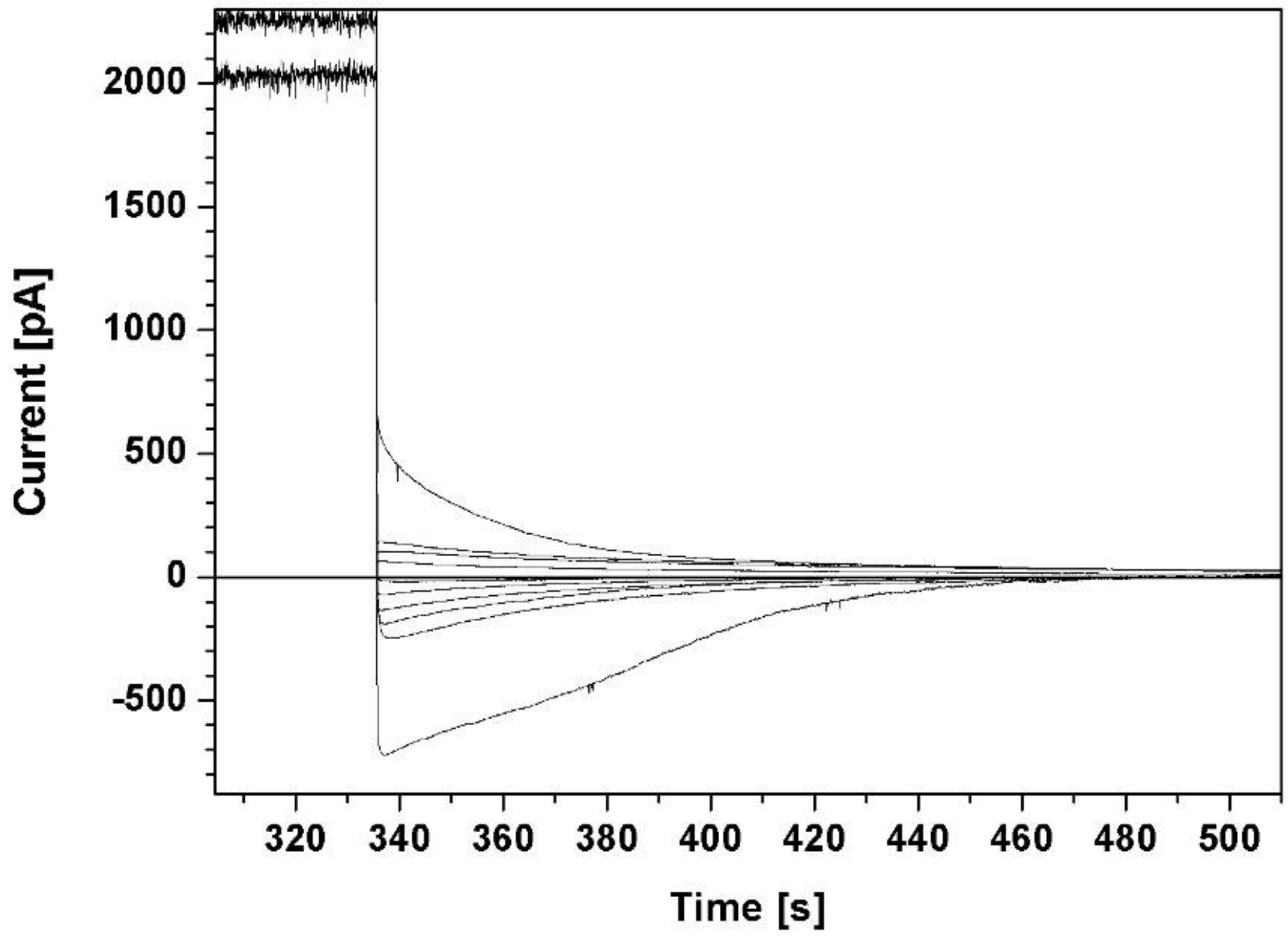

Fig. 12: Response of the CdZnTe sensor to the beam cut-off. Typical pixels were selected for this representation.

### 4.2 Imaging Examples

The imaging performance and the quality of the extractable energy information are demonstrated in figures 13 and 14. The limitation imposed by the small number of pixels in the current prototype assembly was overcome by scanning the object with an x/y-linear stage. A first picture that has been acquired with this setup using a CdZnTe sensor is shown in figure 13. The three pictures of the paper clip show the information of the photon counter, the integrator and the derived mean energy of the detected photon spectrum per pixel. While there is no significant difference in the counter and integrator images, the combined image reveals the increased beam hardening in the area where two wires overlap. This can be seen from the brighter area in the bottom left part of the clip. The integrator image has been flat-field corrected whereas the photon counter underwent no correction except for the usual threshold adjustment. Due to the already mentioned count rate and sensitivity dispersion, the combined image exhibits a visible fixed-pattern-noise.

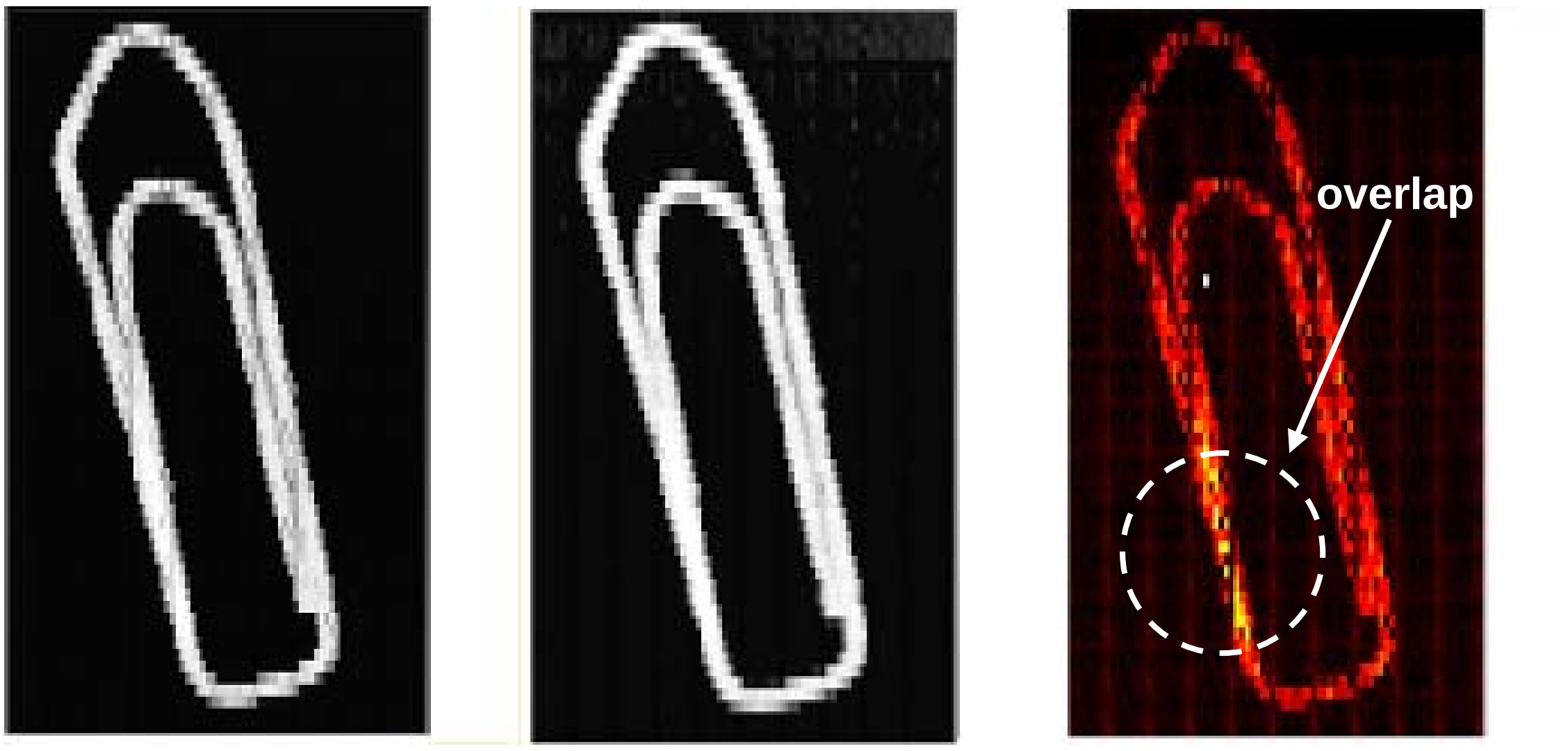

Fig. 13: Radiograph of a paper clip: photon counter image (left), integrator image (middle) and calculated mean energy of the detected spectrum (right). The image was scanned with an 8 x 8 CdZnTe pixel sensor.

The influence of the inhomogeneities has been eliminated in the measurement displayed in figure 14 by setting the scan step size to one pixel dimension and computing the average for all corresponding measurements. While this certainly idealizes the sensor performance, the measurement gives a good illustration of the spectral enhanced imaging capabilities. Again, the counter and the integrator views show no obvious difference in contrast. The image of the reconstructed mean energy on the other hand, shows a strong contrast in the beam hardening in the upper part of the tooth. While this demonstrates the potential benefit of the mean energy reconstruction for contrast enhancement, there is a discrepancy between the expected and measured energy value: the expected mean energy in the direct beam is 33 keV compared to a measured value of 50 keV. The reason for this difference is not fully understood and currently under investigation.

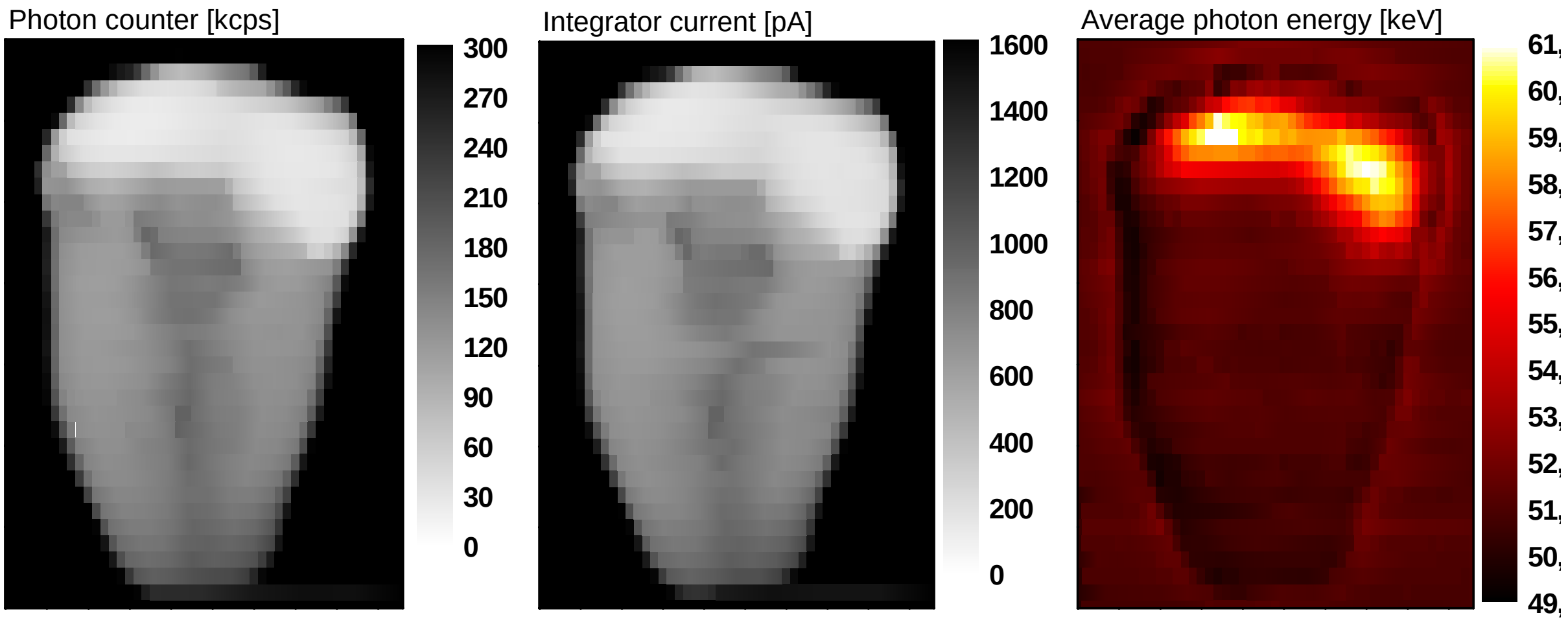

Fig. 14: Radiograph of a tooth (CdZnTe, 90 kVp, 300µm Al filter, 10 keV counter threshold). Scan step size: single pixel.

## 5. CONCLUSION

The counting and integrating signal processing scheme was developed to access additional spectral information from detected X-ray quanta and to increase the dynamic range of the detector system beyond the saturation limit of a photon counter. The second generation of the prototype chip (CIX 0.2) was implemented with a matrix of 8 x 8 pixels and a pixel size of 250 µm x 500 µm. The counter and the integrator show an excellent performance. Simultaneous operation allows a reconstruction of the mean energy in a range of two decades. The use of low noise digital logic permits a dead-time-free readout of the chip with a frame rate up to 20 kHz without compromising the analog performance. CdTe and CdZnTe crystals were bump bonded to the CIX chip and the quality of the imaging properties have been evaluated. The proof of concept for the mean energy reconstruction has been shown. However, the homogeneity and dynamic properties of the sensors limit the imaging performance and the accuracy of the reconstructed mean energy information.

## 6. ACKNOWLEDGEMENTS

The presented work was conducted in a joint research project with Philips Research Europe. The author would like to thank his colleagues Edgar Kraft and Johannes Fink for providing the measurements presented in this paper.